\def\mytitle{Dual-Quaternion Interpolation}

\def\mykeywords{dual-quaternion, quaternion, interpolation, control, graphics, animation}

\def\mysubtitle{ }

\def\myauthor{ Benjamin Kenwright }
\def\myemail{ bkenwright@ieee.org }

\def\myabstract
{
\noindent % Dual-quaternions are a powerful tool for representing position and transitional information in a unified form.
Transformations in the field of computer graphics and geometry are one of the most important concepts for efficient manipulation and control of objects in 2-dimensional and 3-dimensional space. 
Transformations take many forms each with their advantages and disadvantages.
A particularly powerful tool for representing transforms in a unified form are dual-quaternions.
A benefit of this unified form is the interpolation properties, which address a range of limitations (compact form that allows a rotational and translational components to be coupled).
In this article, we examine various dual-quaternion interpolation options that achieve different trade-offs between computational cost, aesthetic factors and coupling dependency.
Surprisingly, despite dual-quaternions being a common tool in graphics libraries, there are limited details on the interpolation details. 
Here we attempt to explain interpolation concept, elaborating on underpinning theories, while explaining concepts and bespoke modifications for added control.

}

\documentclass[9pt, journal,cspaper,compsoc,onecolumn]{IEEEtran}

\usepackage{times}

\usepackage{graphicx}
\usepackage{watermark}

\usepackage[backref=page]{hyperref}

\hypersetup{
    colorlinks = true, 
    linkcolor = blue,
    anchorcolor = red,
    citecolor = blue, 
    filecolor = red, 
}

\usepackage{tikz}
\usetikzlibrary{shapes,arrows}

\usepackage{pgf-pie}

\usepackage{tikz}
\usepackage{calc}% http://ctan.org/pkg/calc
\usepackage{setspace}

\usetikzlibrary{mindmap}

\usepackage{pgfplots}

\usepackage{pgf-pie}

\usepackage{smartdiagram}

\usepackage{bchart}

\usepackage{subcaption}

\usepackage{graphicx}
\graphicspath{{./images/}}

\newcommand{\figureW}[4]{
	\begin{figure}[htbp]
		\centering
		\includegraphics[width=#4\columnwidth]{#1}
		\caption[#2]{\textbf{#2} - #3}
		\label{fig:#1}
	\end{figure}
}

\usepackage{color}  
\usepackage{xcolor}
\definecolor{lbcolor}{rgb}{0.98,0.98,0.98}

\usepackage{listings}

\newcommand{\subparagraph}{}

\usepackage{titlesec}

\renewcommand\theparagraph{}

\titleformat*{\paragraph}{\bfseries}
\titleformat{\paragraph}[runin]
	{\normalfont\normalsize\bfseries \fontsize{9}{9}\selectfont}
	{\theparagraph}
	{0em}
	{}

\titlespacing*{\paragraph}
{0pt}
{1.25ex plus 1ex minus 1.2ex}
{1em}

\usepackage{array}
\usepackage{pifont}

\usepackage{colortbl}

\usepackage{tocloft}

\usepackage{amsmath} % alignedat

\usepackage{amssymb}
\usepackage{amsmath}

\let\labelindent\relax
\usepackage{enumitem}
\setlist[itemize]{noitemsep,leftmargin=*}
\setlist[enumerate]{noitemsep,leftmargin=*}

\begin{document}

\raggedbottom
\sloppy

\title{\fontsize{16}{16}\selectfont \mytitle \\ \fontsize{13}{20}\selectfont \mysubtitle}

%------------------------------------------------------------

% \pdfminorversion=4

\hypersetup{pdfinfo={
   Author		= {\myauthor},
   Title		= {\mytitle  \mysubtitle},
   Subject 		= {\mytitle  \mysubtitle},
   CreationDate = {D:20120220195600},
   Keywords 	= {\mykeywords},
}}

\author{\myauthor \\
{%
%\small 
%School of Computing and Mathematics \break
%Heriot-Watt University, Edinburgh, United Kingdom 
}
\IEEEcompsocitemizethanks{\IEEEcompsocthanksitem \myauthor \protect\\
\myemail % E-mail: bkenwright@ieee.org
}% <-this % stops a space
\thanks{}}

\raggedbottom

% \markboth{Journal of Visualization and Computer Graphics}%

% \markboth{Computer Graphics \& Animation}%
%\markboth{IEEE Transactions on Games (\myauthor)}%
%{Technical Paper}
\markboth{Article (\myauthor)}%
{Educational Paper}

% PRIOR to the title within the \IEEEcompsoctitleabstractindextext IEEEtran
\IEEEcompsoctitleabstractindextext{%
\begin{abstract}
\boldmath
%\fontsize{8}{8}\selectfont
%\normalfont\normalsize\bfseries \fontsize{9}{9.5}\selectfont
%
%
\myabstract

\end{abstract}

}

% ---------------------------------------------------------------------------

%
%%\usepackage[outermarks]{titlesec}
%%\titleformat{\paragraph}
%%  {\normalfont\normalsize\bfseries}{\theparagraph}{1em}{}
%% 
%%\titlespacing*{\paragraph}    {0pt}{3.25ex plus 1ex minus .2ex}{1.5ex plus .2ex}
%
%\setlength\parindent{0pt}
%
%\usepackage{titlesec}
%
%\renewcommand\theparagraph{}
%
%%ref: http://www.ctex.org/documents/packages/layout/titlesec.pdf
%%\usepackage{titlesec}
%%\titleformat{\paragraph}
%%    [runin]
%%    {\bfseries}
%%    {\theparagraph.} 
%%    {1em}
%%    {}
%%    []
%%\titleformat{\paragraph}[hang]{\bfseries}{\theparagraph}{1em}{}
%%\titlespacing*{\paragraph}{0pt}{1.25ex plus 0.5ex minus 0.25 ex}{0.5ex}
%
%%\titlespacing*{\paragraph}{0pt}{1.25ex plus 0.5ex minus 0.25 ex}{0.5ex}
%
%%\titleformat*{\paragraph}{\bfseries}
%\titleformat{\paragraph}[runin]
%	{\normalfont\normalsize\bfseries}
%	{\theparagraph}
%	{0em}
%	{}
%
%%\titlespacing*{\paragraph}
%%{0pt}
%%{3.25ex plus 1ex minus .2ex}
%%{1em}
%
%%ref: http://tex.stackexchange.com/questions/4637/correct-use-of-paragraph-titles/4646#4646
%%\let\originalparagraph\paragraph
%%\renewcommand{\paragraph}[2][:]{\originalparagraph{#2#1}}
%
%
%% ++++++++++++++++++++++++++++++++++++++++++++++++++++++++++++++++++++++++++
%
%\hypersetup{
%    colorlinks 	= true, 
%    linkcolor 	= blue,
%    anchorcolor = red,
%    citecolor 	= blue, 
%    filecolor 	= red, 
%    %pagecolor 	= red,
%    urlcolor 	= red,
%}
%

%%\usepackage{array}

% when
%  who
%   where
%    what
%     why
%      how

\maketitle

\IEEEdisplaynotcompsoctitleabstractindextext
\IEEEpeerreviewmaketitle

\setlength\cftparskip{-2pt}
\setlength\cftbeforesecskip{1pt}
\setlength\cftaftertoctitleskip{2pt}
% \tableofcontents

%-------------------------------------------------------------------------

%% ------------------------------------------------

\section{Introduction}
Dual-quaternions are now a well-established tool in computer graphics for representing rotational and translation solutions in a unified form \cite{kenwrightgraphicssurvey}.
Applications have been explored in a range of areas from
articulated rigid body structures \cite{kenwright2012beginners} and
inverse kinematic systems \cite{kenwrigth2013inverse} to
skinning \cite{kavan2007skinning} and
surfaces and curves \cite{kensurfacesandcurves}.

\paragraph{Contribution}
The key contributions of this article are the explanation and application of dual-quaternions specifically in and around interpolation.
We review and present dual-quaternion interpolation techniques in addition to a hybrid approach (called KenLERP) to control the coupled connection between the rotation and translational components for aesthetic qualities.
We provide practical details in addition to an interactive example so readers can visually assess and compare the different interpolation techniques.

\section{Dual-Quaternion Conversion Operations}
To keep the article complete and to provide a consistent set of notation criteria for conversion between standard types, we present the fundamental conversion equations here.

%DQ:
%ζ= q_r+ q_d ϵ

\paragraph{Translation and Rotation to Dual-Quaternion}
Given a quaternion rotation ($q$) and a three dimensional translational  vector ($\vec{t}$) the dual-quaternion ($\zeta(q,\vec{t})$) is given below:

\begin{equation}
\begin{alignedat}{5}
     \zeta(q,\vec{t}) &= q_r + q_d \\
                 q_r  &= q \\
                 q_t  &= (0,t_x,t_y,t_z)  \hspace{10pt} \text{quaternion, w=0} \\
                 q_d  &= ( q_r q_t ) \frac{1}{2}
\end{alignedat}
\end{equation}

\paragraph{Dual-Quaternion to Position and Rotation}

Given a dual-quaternion ($\zeta$) extracting the quaternion rotation ($q$) and translation vector ($\vec{t}$) components is given below:

\begin{equation}
\begin{alignedat}{5}
       \zeta &= q_r + q_d \\
       q     &= q_r \\
       q_t   &= (2 q_d) q_r^* \\
     \vec{t} &= (q_{tx}, q_{ty}, q_{tz} ) 
\end{alignedat}
\end{equation}

While the two main forms above show the conversion to and from the vector forms it should be noted that it is just as straightforward and simple to convert to and from the matrix version as shown by Kenwright \cite{kenwright2012beginners}.

\section{Dual-Quaternion Interpolation Types}

We explain, compare and demonstrate four main ways of interpolating dual-quaternions, including the uncomplicated `decoupled' approach and screw linear interpolation through to a hybrid based approach for aesthetic control (damping or limiting the coupling factor).

% Dual-Quaternion Interpolation Types 
\paragraph{SEP(LERP)}
SEP(LERP) – extract translation and orientation, interpolate separately and reconstruct new dual-quaternion.

\begin{equation}
\begin{alignedat}{5}
  \text{LERP}(t;\vec{v0},\vec{v1}) &= (1-t)\vec{v0} + (t)\vec{v1} \\
  \text{SLERP}(t;q_0, q_1)         &= q_0 (q_0^* q_1)^t
\end{alignedat}
\end{equation}

\paragraph{DLB} Dual quaternion Linear Blending – interpolate each component (8 separate floats), renormalized, a generalization of
QLB \cite{kavan2005spherical}.

\begin{equation}
\text{DLB}(t;\zeta_A,\zeta_B) = \frac{(1-t)\zeta_A + (t\zeta_B)}{||(1-t)\zeta_A + (t\zeta_B)||}
\end{equation}

\paragraph{ScLERP} – dual-quaternion power/screw parameters
ScLERP (Screw Linear Interpolation), a generalization of
SLERP \cite{shoemake1985animating}.

\begin{equation}
\text{ScLERP}(t;\zeta_A,\zeta_B) = \zeta_A (\zeta_B^*  \zeta_B)^t
\end{equation}

\paragraph{KenLERP} 
Hybrid approach to control the influence between the rotational and translational component during interpolation.
Add a bias factor $\beta$ between 0.0 and 1.0. 
A value of 0.0 decouples the rotation and translation components while a value of 1.0 creates a fully coupled response (ScLERP). 
There is also the condition for values greater than 1.0, which adds an `amplification' factor for exaggerating interpolation values (for aesthetic impact - comparable with over/under damping).

%(SEP and ScLERP) – coupling between rot and trans

\begin{equation}
\begin{alignedat}{5}
  [v_0,q_0] = ScLERP(\zeta_A, \zeta_B, t) \\
  [v_1,q_1] = SEPLERP(\zeta_A, \zeta_B, t)\\
  \zeta = [LERP(v_0,v1,\beta), SLERP(q0,q1,\beta)]
\end{alignedat}
\end{equation}

The hybrid approach allows the interpolation to be controlled, % Controlling the Interpolatio
various factors that affect the aesthetic quality. %  of the interpolant
See Ken2020 \cite{kenlerp2020} for an interactive prototype of the concept.

%% ------------------------------------------------

\section{General Discussion and Conclusion}

The advantages of being able to control the coupled nature of dual-quaternions during interpolation has advantages in graphical contexts for controlling the aesthetic qualities of the interpolation values.
We have demonstrated the viability and practicality of a hybrid based approach.

% ---------------

\appendix

\section{Appendix}

\subsection*{Dual-Quaternion Algebra}
For completeness, we include the mathematical details for essential dual-quaternion operations.

\paragraph{Dual-Quaternion Arithmetic Operations}

The elementary arithmetic operations necessary for us to use dual-quaternions are given below:

\begin{itemize}

\item \textbf{dual-quaternion}: $\zeta = q_r + q_d \varepsilon$

\item \textbf{scalar multiplication}:  $s \zeta = s q_r + s q_d \varepsilon$

\item \textbf{addition}: $\zeta_1 + \zeta_2 = q_{r1} + q_{r2} + (q_{d1}+q_{d2}) \varepsilon$

\item \textbf{multiplication}: $\zeta_1 \zeta_2 = q_{r1} q_{r2} + (q_{r1}q_{d2} + q_{d1}q_{r2}) \varepsilon$

\item \textbf{conjugate}:  $\zeta^* = q_r^* + q_d^* \varepsilon$

\item \textbf{magnitude}: $||\zeta|| = \zeta \zeta^{*}$

\end{itemize}
\noindent where $q_r$ and $q_d$ indicate the real and dual part of a dual-quaternion (we use a `$q$' to emphasis that they are quaternions).

For a beginners introduction to dual-quaternions and a comparison of alternative methods (e.g., matrices and Euler angles) and how to go about implementing a straightforward library we refer the reader to the paper by Kenwright \cite{kenwright2012beginners}.

\paragraph{Dual-Quaternion Vector Transformation}
A dual-quaternion is able to transform a 3D vector coordinate as shown in Equation \ref{eq:dqtransf}. Note that for a unit dual-quaternion the inverse is the same as the conjugate \footnote{
Just to note, there are three definitions for the conjugate of a dual quaternion: 
1: $\zeta^* = q_r^* + q_d^* \varepsilon$,
2: $\zeta^* = q_r - q_d  \varepsilon$, and
3: $\zeta^* = q_r^* - q_d^* \varepsilon$.
For transforming a point using Equation \ref{eq:dqtransf}, you would use the 2nd conjugate variation.
% ref: https://www.euclideanspace.com/maths/algebra/realNormedAlgebra/other/dualQuaternion/functions/index.htm
}.

\begin{equation}
    p' = \hat{\zeta} p \hat{\zeta}^{-1}
    \label{eq:dqtransf}
\end{equation}

\noindent where $\hat{\zeta}$ is a unit dual-quaternion representing the transform, $\hat{\zeta}^{-1}$ is the inverse of the unit dual-quaternion transform.
$p$ and $p'$ are the dual-quaternions holding 3D vector coordinate to before and after the transformation (i.e., $p=(1,0,0,0) + \epsilon (0,v_x,v_y,v_z)$ )).

\paragraph{{Pl{\"u}cker Coordinates}}
Pl{\"u}cker coordinates are used to create Screw coordinates which are an essential technique of representing lines. We need the Screw coordinates so that we can re-write dual-quaternions in a more elegant form to aid us in \textbf{formulating a neater and less complex interpolation method} that is comparable with spherical linear interpolation for classical quaternions.

The Definition of Plu\"cker Coordinates:
\begin{itemize}
\item $\vec{p}$ is a point anywhere on a given line
\item $\vec{l}$ is the direction vector
\item $\vec{m} = \vec{p} \times \vec{l}$ is the moment vector
\item $(\vec{l},\vec{m})$ are the six Plu\"cker coordinate
\end{itemize}

We can convert the eight dual-quaternion parameters to an equivalent set of eight screw coordinates and vice-versa.
The definition of the parameters are given below in Equation \ref{eq:dqscrew}:

\begin{equation}
\begin{alignedat}{4}
\text{screw parameters} &= (\theta,d,\vec{l},\vec{m}) \\
\text{dual-quaternion}  &= q_r + \epsilon q_d \\
                        &= (w_r+\vec{v}_r) + \epsilon(w_d + \vec{v}_d) 
\end{alignedat}
\label{eq:dqscrew}
\end{equation}

\noindent where in addition to $\vec{l}$ representing the vector line direction and $\vec{m}$ the line moment, we also have $d$
representing the translation along the axis (i.e., pitch) and the angle of rotation $\theta$.

\figureW
{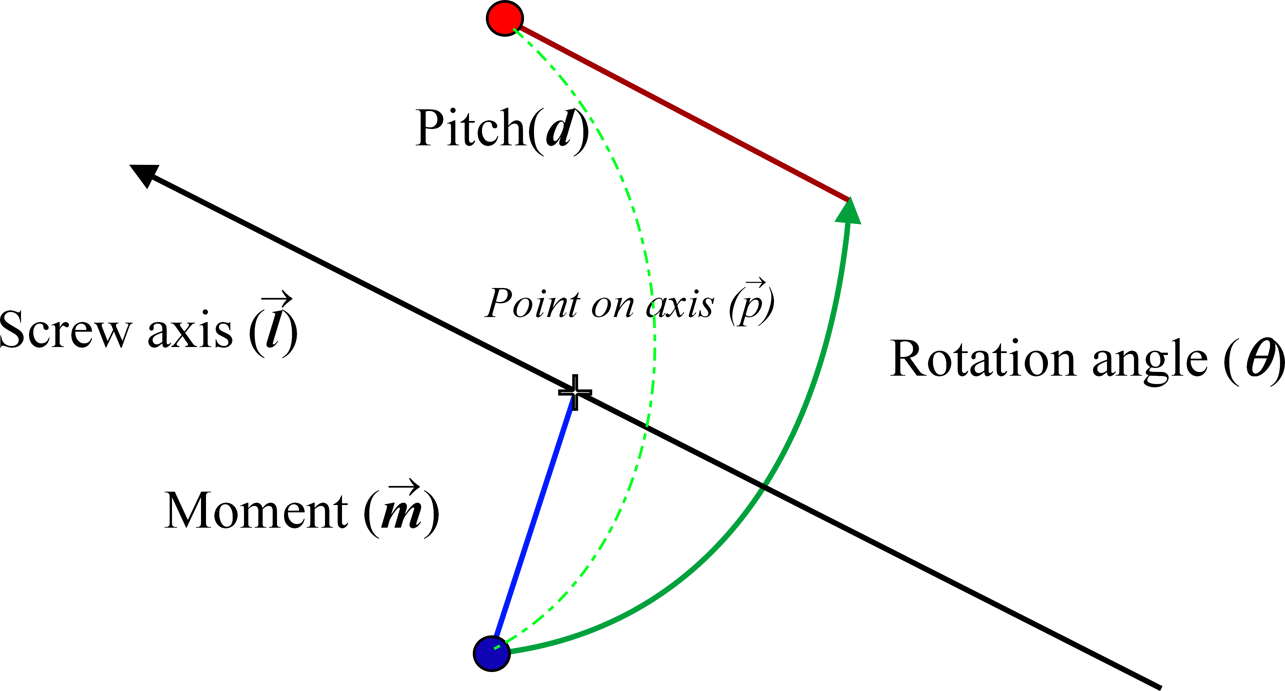}
{SCREW Parameters}
{The displacement for a dual-quaternion is equivalent to a rotation about some screw axis followed by a translation along the axis.}
{0.5}

Daniilidis  \cite{daniilidis1999hand} provided a breakdown of the screw parameters in relation to a dual-quaternion (given below in Equation \ref{eq:dquatandscrew}).
\begin{equation}
\begin{alignedat}{3}
\hat{\zeta} &= \left(
                 w_r
                 \atop
                 \vec{v}_r
               \right)
                 + \epsilon
               \left(
                 w_d
                 \atop 
                 \vec{v}_d
               \right) \\
            &= \left(
                 w_r
                 \atop 
                 \vec{v}_r
               \right)
                 + \epsilon
               \left(
                  -\frac{1}{2}\vec{v}_r \cdot \vec{t}
                  \atop
                  \frac{1}{2}(w_r \vec{t} + (\vec{t} \times \vec{v}_r))
               \right)\\
            &= \left(
                     cos(\frac{\theta}{2})
                     \atop
                     sin(\frac{\theta}{2}) \vec{l}
               \right)
               + \epsilon
               \left(
                    {-\frac{d}{2}sin(\frac{\theta}{2})}
                    \atop
                    {sin(\frac{\theta}{2})\vec{m} +
                    \frac{d}{2}
                    cos(\frac{\theta}{2})
                    \vec{l}}
               \right) 
\end{alignedat}
\label{eq:dquatandscrew}
\end{equation}

\textbf{Convert screw-parameters to dual-quaternion}

Given Equation \ref{eq:dquatandscrew}, we can derive the associated dual-quaternion components in relation to specific screw-parameters (and vice versa).

\begin{equation}
\begin{alignedat}{3}
w_r &= cos \left( \frac{\theta}{2} \right) \\
\vec{v}_r &= \vec{l} sin \left( \frac{\theta}{2} \right) \\
w_d &= -\frac{d}{2} sin \left( \frac{\theta}{2} \right) \\
\vec{v}_d &= sin \left( \frac{\theta}{2} \right) \vec{m} + \frac{d}{2} cos \left( \frac{\theta}{2} \right) \vec{l}
\end{alignedat}
\end{equation}

\textbf{Convert dual-quaternion to screw-parameters}

\begin{equation}
\begin{alignedat}{3}
\theta    &= 2 cos^{-1}(w_r)   \hspace{25pt} && \text{(angle from real component)}\\
\vec{l}   &= \vec{v}_r \frac{1}{sin(\frac{\theta}{2})}  \hspace{20pt} && \text{(or $||v_r|||$ axis from the real component)}\\
     d    &= -w_d \frac{2}{sin(\frac{\theta}{2})}        \hspace{20pt} && \text{(or $\vec{t} \cdot \vec{l}$ )} \\
\vec{m} &= v_d \frac{1}{sin(\frac{\theta}{2})} - \frac{d}{2}\frac{cos(\frac{\theta}{2})}{sin(\frac{\theta}{2})}\vec{l}  \hspace{20pt} && \text{(cos/sin = cot)}\\
\end{alignedat}
\end{equation}

\iffalse
\begin{equation}
\begin{alignedat}{3}
\theta &= 2 cos^{-1}(w_r) \\
d      &= -2w_d \frac{1}{\sqrt(\vec{v}_r \dot \vec{v}_r)} \\
\vec{l} &= \vec{v}_r \left( \frac{1}{\sqrt(\vec{v}_r \dot \vec{v}_r)} \right) \\
\vec{m} &= \left( \vec{v}_d - \vec{l} \frac{d w_r}{2} \right) \frac{1}{\sqrt(\vec{v}_r \dot \vec{v}_r)}
\end{alignedat}
\end{equation}
\fi

\paragraph{Dual-Quaternion power}

We can write the dual-quaternion representation in the form given in Equation \ref{eq:dqpow}. %  (see Daniilidis [14] for details).

\begin{equation}
\begin{alignedat}{3}
\hat{\zeta} &= cos \left( \frac{\theta + \epsilon d}{2} \right)
            + (\vec{l} + \epsilon \vec{m}) 
               sin \left( \frac{\theta + \epsilon d}{2} \right) \\
            &= cos \left( \frac{\hat{\theta}}{2} \right) +
                \hat{v} \sin \left( \frac{\hat{\theta}}{2} \right)
\end{alignedat}
\label{eq:dqpow}
\end{equation}

\noindent where $\hat{\zeta}$ is a unit dual-quaternion, $\hat{v}$ is a unit dual-vector ($\hat{v}=\vec{l}+\epsilon\vec{m}$), and $\hat{\theta}$ is a dual-angle ($\hat{\theta}=\theta+\epsilon d$).

The dual-quaternion in this form is exceptionally
interesting and valuable as it allows us to calculate a dual-quaternion to a power. Calculating a dual-quaternion to a power is essential for us to be able to easily calculate
spherical linear interpolation. However, \textbf{instead of purely rotation as with classical quaternions, we are instead now able to interpolate full 6-dimensional degrees of freedom} (i.e., rotation and translation) by using dual-quaternions. 

\begin{equation}
    \hat{\zeta}^{t} = cos \left( t \frac{\hat{\theta}}{2} \right) + 
    \hat{v} sin \left( t \frac{\hat{\theta}}{2} \right)
\end{equation}

\paragraph{Dual-Quaternion Screw Linear Interpolation (ScLERP)}
ScLERP is an extension of the quaternion SLERP
technique, and allows us to create constant smooth interpolation between dual-quaternions. 
Similar to quaternion SLERP, we use the power function to calculate the interpolation values for ScLERP shown in Equation \ref{eq:dqsclearp}.

\begin{equation}
\begin{alignedat}{3}
\text{ScLERP}(\hat{\zeta}_A, \hat{\zeta}_B : t ) = \hat{\zeta}_A (\hat{\zeta}^{-1}_A \hat{\zeta}_B )^t
\end{alignedat}
\label{eq:dqsclearp}
\end{equation}

\noindent where
$\hat{\zeta}_A$
and
$\hat{\zeta}_B$
are the start and end unit dual-quaternion and $t$ is the interpolation amount from $0.0$ to $1.0$.

Alternatively, a \textbf{fast approximate alternative to ScLERP} was presented by Kavan et al. \cite{kavan2007skinning} called Dual-Quaternion Linear Blending (DLB). Furthermore, dual-quaternions have gained a great deal of attention in the area of character-based skinning. Since, a skinned surface approximation using a weighted dual-quaternion approach produces less kinking and reduced visual anomalies compared to linear methods by ensuring the surface keeps its volume.

\bibliographystyle{plain}

%\bibliographystyle{ieeetran}

% \bibliographystyle{ieeetr}

% \bibliographystyle{ACM-Reference-Format}

%\fontsize{8.0}{8.0}\selectfont

% \vspace{-15pt}

\bibliography{paper} % bib filename

\end{document}